\def\rf#1;#2;#3;#4;#5 {\par#1, {\it #3} {\bf #4}, #5 (#2). \par}

\def\beq#1{\begin{equation}\label{#1}}
\def\eeq{\end{equation}}
\def\beqa#1{\begin{eqnarray}\label{#1}}
\def\eeqa{\end{eqnarray}}

\def\ignore#1{}

\def\spose#1{\hbox to 0pt{#1\hss}}
\def\simlt{\mathrel{\spose{\lower 3pt\hbox{$\mathchar"218$}}
     \raise 2.0pt\hbox{$\mathchar"13C$}}}
\def\simgt{\mathrel{\spose{\lower 3pt\hbox{$\mathchar"218$}}
     \raise 2.0pt\hbox{$\mathchar"13E$}}}
\def\simpropto{\mathrel{\spose{\lower 3pt\hbox{$\mathchar"218$}}
     \raise 2.0pt\hbox{$\propto$}}}

\def\ed{\end{document}}
\def\bl{\vskip0.423truecm}
\def\ind{\noindent\hskip1.8truecm}
\def\Section#1{{\raggedright\bl\bl\goodbreak{\noindent\bf#1}\bl}}

\documentclass [11pt]{article}

\begin{document}

{\hfill} \vskip1.222truecm


\noindent
{\bf THE THERMODYNAMICAL ARROW OF TIME: REINTERPRETING
THE BOLTZMANN-SCHUETZ ARGUMENT}

\bl\bl\bl \ind Milan M. \'Cirkovi\'c

\bl\ind {\it Astronomical Observatory Belgrade

}

\ind {\it Volgina 7, 11160 Belgrade}

\ind {\it Serbia and Montenegro}

\ind {\it arioch@eunet.yu}

\bl\bl\bl\ind Received April 25, 2002; revised October 27, 2002

\bl\bl

\baselinestretch

\noindent \textbf{Abstract.} The recent surge of interest in the
origin of the temporal asymmetry of thermodynamical systems
(including the accessible part of the universe itself) put forward
two possible explanatory approaches to this age-old problem.
Hereby we show that there is a third possible alternative, based
on the generalization of the classical (``Boltzmann-Schuetz'')
anthropic fluctuation picture of the origin of the perceived
entropy gradient. This alternative (which we dub the
Acausal-Anthropic approach) is based on accepting Boltzmann's
statistical measure at its face value, and accomodating it within
the quantum cosmological concept of the multiverse. We argue that
conventional objections raised against the Boltzmann-Schuetz view
are less forceful and serious than it is usually assumed.
\textit{A fortiori}, they are incapable of rendering the
generalized theory untenable. On the contrary, this analysis
highlights some of the other advantages of the multiverse approach
to the thermodynamical arrow of time.


\clearpage

\textit{If the doors of perception were cleansed,}

\textit{everything would appear to man as it is, infinite.}

\

William Blake, ``The Marriage of Heaven and Hell'' (1793)

\vspace{0.5cm}


\Section{1. INTRODUCTION: THE NATURE OF THERMODYNAMICAL
EXPLANANDUM}

\noindent
The problem of the time-asymmetry of
thermodynamics---already more than a century old in its
\textit{modern} form!---is the following. In our experience,
systems increase in entropy in the forward direction of time. The
underlying dynamical laws, which are taken to govern thermodynamic
systems, however, are symmetric in time: statistical mechanics
predicts that entropy is overwhelmingly likely to increase in both
temporal directions. So where does the asymmetry of thermodynamics
and of our experience generally come from? It was, of course, the
great Ludwig Boltzmann who---prompted by Loschmidt, Culverwell and
Burbury (joined later by Zermelo)---asked that deep question,
contingent on his statistical explanation of thermodynamical
phenomena; in his words, ``is the apparent irreversibility of all
known natural processes consistent with the idea that all natural
events are possible without restriction?''$^{(\ref{eq1})}$

During the last decade we have been witnessing a renaissance of interest in
the problem of thermodynamical asymmetry of the world around us among
physicists and philosophers alike (e.g., Refs. 2-7). In two recent
remarkably clear and interesting papers, Australian philosopher of science
Huw Price attempted to show that there are two competing projects for the
explanation of the perceived thermodynamical asymmetry, which he labels
Causal-General and Acausal-Particular approaches.$^{(8,9)}$ Furthermore, his
intention was to show the superiority of the Acausal-Particular approach,
which is in accordance with other pieces of his atemporal worldview, which
he also presented in his recent brilliant monograph on the subject of
temporal asymmetries (Ref. 3). The answer Price advocates relies on a
low-entropy initial boundary condition: if the initial state of the universe
is one of extremely low entropy, then Boltzmannian statistical
considerations yield an overwhelmingly likely entropy increase towards the
future throughout the history of the universe. Price contrasts this
account---a version of what he calls the \textit{Acausal-Particular} approach---with those
theories that derive thermodynamic asymmetry from some underlying asymmetric
causal or dynamical mechanism operating at all times (like the
quantum-mechanical state reduction in quantum theories with dynamical
reduction), what he calls \textit{Causal-General} views (cf. Ref. 5). Causal-General views
necessarily contradict Boltzmann's attitude toward the time-reversal
asymmetry: ``This one-sidedness lies uniquely and solely in the initial
conditions.'' To Price's mind, these two kinds of account are the only
serious contenders to explain the time-asymmetry of thermodynamics.

In this paper, we would like to show that this account (implicitly accepted
by other recent writers on the topic) is biased and lacking some important
ingredients. While the desire for clarification of the common explanatory
task is highly commendable, it is important that the taxonomy is kept both
maximally comprehensive \textit{and} just. The suggested taxonomy fails in one important
respect: it fails to notice an alternative to both the Causal-General and
Acausal-Particular views. Therefore, we would like here to point out that
from Price's Acausal-Particular approach bifurcates another option which
deserves a separate mention in reviewing the ways toward the explanation of
thermodynamical asymmetry. This third approach differs markedly from the
other two in its conception of what needs to be done to solve the puzzle. In
proposing this, we follow the lead of Price himself who, introducing his two
proposals, points out that (Ref. 9, sec. 1.1)

\begin{quotation}

So far as I know, the distinction between these two approaches has not been
drawn explicitly by other writers. Without it, it is not easy to appreciate
the possibility that many familiar attempts to explain the time-asymmetry of
thermodynamics might be not \textit{mistaken} so much as \textit{misconceived}---addressed to the wrong problem,
in looking for time-asymmetry in the wrong place.

\end{quotation}

We shall see that, unfortunately, ``those who use the sword will be killed
by the sword'', and that there are several instances in which Price's own
favorite proposal is misconceived, for \textit{exactly} the same reason: searching for a
solution to the puzzle in the wrong place.

The second goal of the present note is to demonstrate that the classical
arguments (accepted unquestioningly by Zeh, Price, Albert and other
contemporary authors) against the classical Boltzmann-Schuetz anthropic
fluctuation picture are much less forceful than it was hitherto assumed.
This serves an important role in bringing the atypical initial cosmological
conditions within the domain of physical explanation.


\Section{2. ACAUSAL-ANTHROPIC APPROACH}

\noindent
Our first motivation is a full and faithful acceptance
of Price's account of the nature of the explanatory task ahead
(Ref. 8, section 2.2):

\begin{quotation}

First of all, let's assume that basic explanatory questions are of the form:
'Given that C, why E rather than F?' The thought here is that we never
explain things in isolation. We always take something as already given, and
seek to explain the target phenomena in the light of that. C represents this
background, and E the target phenomenon. (C might comprise accepted laws, as
well as 'boundary conditions' being treated as 'given' and unproblematic for
the purposes at hand.)

\end{quotation}

We should, naturally, seek to make C maximally comprehensive.
Obviously, our existence as observers is part of the necessary
background. Should it not be included in C? (``It is very much in
the spirit of scientific inquiry to welcome any theory which
widens the range of applicability of science.''---Ref.\ 10, p.\
74) In most practical physical applications it is irrelevant, at
least in the first approximation (and therefore is usually
omitted, although reasons for the omission are seldom explicated).
Our existence as observers hardly impacts on the explanatory
project for, say, matter-enhanced neutrino oscillations or
fractional Hall effect. However, \textit{in the}
\textit{cosmological context}, leaving observers out of the
picture does not lead to happy consequences, as was first shown by
Dicke and Carter in 1960s and 1970s. The debate between Dirac and
Dicke on the nature of explanatory projects for the ``large number
coincidences'' is especially instructive in this respect, since
several parallels with the projects for explaining thermodynamical
asymmetry can be drawn. The reader may recall that the famous
``Dirac's hypothesis'' (or the ``large-number hypothesis'') to
explain the ubiquitous dimensionless number 10$^{40}$, came from a
bold suggestion that all these numerical coincidences are in fact
exact equations related to the age of the universe.$^{(11)}$ In
other words, the relevant number is large because the universe is
old.

{}From this profound and simple (but wrong!) idea, Dirac and later
investigators (e.g., Ref. 12) have deduced a wealth of extremely interesting
and challenging empirical predictions, the most famous being the decrease in
the Newtonian gravitational constant $G$ during cosmological evolution. Using
the completely opposite approach, Dicke$^{(13)}$ has suggested that large
number coincidences are \textit{observed} only because any conceivably different values to
such dimensionless quantities would be incompatible with our existence as
intelligent observers (and consequently the relationships are only
approximate). Consequently, the two explanatory projects, Dirac's and
Dicke's, found themselves in the open arena of physical investigations.

Not only has subsequent empirical evidence disproved Dirac's ideas; notably,
the decreasing $G$ has been spectacularly falsified by experiments with the
\textit{Viking} space probes, as well as in the binary pulsar experiments,$^{(14,15)}$ but
more important is the fact that the subsequent developments show Dirac's
general approach to be rather sterile in comparison to Dicke's. What we are
really interested in here is a comparison of explanatory approaches. We can
schematically compare the view of the large-numbers explanandum in this
case:

\vspace{0.4cm}

\textbf{Dirac}: ``objective'' coincidences (no properties of observers
included in C);

\textbf{Dicke}: ``observed'' coincidences (properties of observers included
in C).

\vspace{0.4cm}

The outcome of that particular duel should warn us against
dogmatism when cosmological theories are the subject of
inquiry.\footnote{ In spite of the fact that Dirac himself
emphasized the ``optimistic'' nature of his explanation in
comparison: ``On Dicke's assumption habitable planets could exist
only for a limited period of time. With my assumption they could
exist indefinitely in the future and life need never
end.''$^{(16)}$ This goes some steps toward addressing vulgar
objections still heard in some quarters that anthropic
explanations are anthropocentric or even ``cozy'' (e.g., Ref.\
17). }

With this in mind, we propose a novel approach to the explanation
of thermodynamical asymmetry, one which could be labeled (for the
reasons explained below) the \textit{Acausal-Anthropic approach}.
It represents a Dicke-like approach applied to the specific
problem of thermodynamical asymmetry and the nature (entropy-wise)
of the cosmological initial conditions. To understand what this
proposal encompasses, perhaps the best way is to use Price's
``parsing'' the natural phenomena for the different approaches
(Ref.\ 8, \S 2.3; Ref.\ 9, \S 3.1-3.2). Applied to the
Acausal-Anthropic approach, it may look like this:

\vspace{0.5cm}

Symmetric boundary conditions---entropy high in the past

Symmetric default condition---entropy likely to be high (always)

+ Asymmetric observational selection effect

{\_}{\_}{\_}{\_}{\_}{\_}{\_}{\_}{\_}{\_}{\_}{\_}{\_}{\_}{\_}{\_}{\_}{\_}{\_}{\_}{\_}{\_}{\_}{\_}{\_}{\_}{\_}{\_}{\_}{\_}{\_}{\_}{\_}{\_}{\_}{\_}{\_}{\_}{\_}{\_}{\_}{\_}{\_}{\_}{\_}{\_}{\_}{\_}{\_}{\_}{\_}{\_}{\_}

Observed asymmetry

\vspace{0.4cm}

This should be compared with similar parsing schemes for the Causal-General:

\vspace{0.4cm}

Asymmetric boundary condition---entropy low in the past

+ Asymmetric law-like tendency---entropy constrained to increase

{\_}{\_}{\_}{\_}{\_}{\_}{\_}{\_}{\_}{\_}{\_}{\_}{\_}{\_}{\_}{\_}{\_}{\_}{\_}{\_}{\_}{\_}{\_}{\_}{\_}{\_}{\_}{\_}{\_}{\_}{\_}{\_}{\_}{\_}{\_}{\_}{\_}{\_}{\_}{\_}{\_}{\_}{\_}{\_}{\_}{\_}{\_}{\_}{\_}{\_}{\_}{\_}{\_}

Observed asymmetry

\vspace{0.4cm}

\noindent
and Acausal-Particular views$^{(8)}$:

\vspace{0.4cm}

Asymmetric boundary condition---entropy low in the past

+ Symmetric default condition---entropy likely to be high, \textit{ceteris paribus}

{\_}{\_}{\_}{\_}{\_}{\_}{\_}{\_}{\_}{\_}{\_}{\_}{\_}{\_}{\_}{\_}{\_}{\_}{\_}{\_}{\_}{\_}{\_}{\_}{\_}{\_}{\_}{\_}{\_}{\_}{\_}{\_}{\_}{\_}{\_}{\_}{\_}{\_}{\_}{\_}{\_}{\_}{\_}{\_}{\_}{\_}{\_}{\_}{\_}{\_}{\_}{\_}{\_}

Observed asymmetry.

\vspace{0.5cm}

The basic idea of the Acausal-Anthropic approach is that, having
already received from (quantum) cosmology a useful notion of the
multiverse, we could as well employ it in order to account for the
\textit{prima facie} extremely improbable choice of (local)
initial conditions. In other words, we imagine that everything
that exists, for which we shall use the term multiverse,\footnote{
Not to be confused with the multiverse of quantum mechanics
(``Everett's multiverse'' = the totality of wavefunction
branches). Here we refer to the multiverse of quantum cosmology
(``Linde's multiverse'' = set of different cosmological domains,
possibly causally and/or topologically disconnected from ours).
This distinction does \textit{not} preclude the relationship
between the two, however. But that relationship belongs to the
quantum-cosmological domain and is certainly beyond the scope of
the present manuscript.} represents a ``Grand Stage'' for the
unfolding of---among other things---the thermodynamical histories
of chunks of matter. Entropy \textit{in the multiverse\/} is high
almost everywhere and at all times (``almost'' here meaning
``everywhere minus a possible subset of small or zero measure'').
In other words, the multiverse is, formally speaking, in the state
of ``heat death'', as envisaged by classical
thermodynamics.$^{(18,19)}$ Our cosmological domain (``the
universe'') represents a natural fluctuation---presumably of small
or zero measure; but the anthropic selection effect answers the
question ``why do we find ourselves on an upward slope of such a
fluctuation?'' Hence what needs explaining is not that there are
such fluctuations (this is entailed by Boltzmann's statistical
measure); nor the fact that the local initial conditions are of
extremely low probability (this results from a distribution over
all domains); but the fact that we happen to live in such an
atypical region of the ``grand total'' which is almost always in
equilibrium. And that is to be explained by showing why the
observed entropy gradient is necessary for our existence as
intelligent observers.\footnote{ Note that we employ the term
``acausal'' here in the same sense as Price does: the absence of
law-like reason explaining a particular feature of the physical
world. This usage has nothing to do with Lorentz invariance,
superluminal motions and the like! It has been recognized for
quite some time in the philosophy of physics that there are two
rather different meanings attached to the notion of ``causality''.
Causality as a geometrical relation between events on the
spacetime manifold is different from causality as an order
relation between stages of a developing physical process. In
accordance with Horwitz, Arshansky, and Elitzur,$^{(67)}$ we may
call the former \textit{spacetime causality\/} and the latter
\textit{process causality}. The Anthropic-Acausal approach
described here is acausal in the sense of process causality, not
in the sense of spacetime causality.}

{}From the point of view of the present Acausal-Anthropic approach, Boltzmann
(and his assisstant Dr. Shuetz to whom he gave credit for the original idea;
see Ref. 20) was on the right track in proposing what came to be known as
the ``anthropic fluctuation'' picture.$^{(21)}$ The idea was to explain the
local thermodynamical disequilibrium by appealing to the size of the
universe and the conditions necessary for our existence. Remember the famous
words of the Viennese master$^{(20)}$:

\begin{quotation}

If we assume the universe great enough, we can make the probability of one
relatively small part being in any given state (however far from the state
of thermal equilibrium), as great as we please. We can also make the
probability great that, though the whole universe is in thermal equilibrium,
our world is in its present state. It may be said that the world is so far
from thermal equilibrium that we cannot imagine the improbability of such a
state. But can we imagine, on the other side, how small a part of the whole
universe this world is? Assuming the universe great enough, the probability
that such a small part of it as our world should be in its present state, is
no longer small.

\end{quotation}

\noindent In other words, in the Boltzmann-Schuetz view the world
in general is in the state of maximal entropy (``heat death''). We
exist within a fluctuation of low entropy (\textit{without}
reason, i.e. acausally!), which makes our existence possible.
Thus, any acausal version of the explanatory project on even
vaguely ``Boltzmannian'' grounds has to include the anthropic
selection effect. Boltzmann and Schuetz thus, in our present view,
were on the right track, and could not have done better under the
circumstances. What they did lack was the input of modern
cosmology, exemplified by the multiverse
concept.\footnote{Boltzmann understood the difficulties of his
position quite well; there are several examples of his regarding
his cosmological thoughts as remote speculations.} If we summon
such help, we may truly inherit the Boltzmannian project of
setting up the explanatory proposal for the observed entropy
asymmetry.

How many domains are required in order to account for the observed
thermodynamical asymmetry? While the exact answer is difficult to
conceive, we may follow the lead of Penrose and use the
Bekenstein-Hawking formula$^{(22,23)}$ to estimate the
\textit{lower limit\/} on the size of ensemble of domains in which
we expect to find one similar to ours on purely probabilistic
grounds. Obviously, the celebrated Boltzmann formula $S = \ln W$
(in ``natural'' $k = 1$ units we shall use in the entire
discussion) suggests that the required number is $N_{\min } \sim
\exp S_{\max } $, where $S_{\max } $ is the entropy of the state
of maximal probability of the matter in our domain (what would be
traditionally called the entropy of the ``heat death'' state).
However, our domain is limited by our particle horizon at present,
and will be almost certainly limited$^{(24,25)}$ by an
\textit{event horizon}, due to the contribution $\Omega _{\Lambda
}$ of the vacuum energy density (``cosmological constant'').
Numerically, the difference between the two in the realistic case
is not very large in cosmological terms ($\sim $1 Gpc), so we will
not make a big error in attributing the state of low entropy to
those currently invisible (but visible to our descendants!) parts
of our domain between the particle and the event horizon. Thus, we
need to account for the entropy of matter of cosmological density
$\Omega _{m}$ (predominantly in CDM or similar particles, with
$\sim $15-20{\%} of baryons). On the assumption that our domain is
globally flat, with no net electric charge and no net angular
momentum, we obtain:

\begin{equation}
\label{eq1}
\begin{array}{l}
 N_{\min } \sim \frac{\exp S_{\max } }{\exp S_0 } = \frac{1}{\exp S_0 }\exp
\left( {\frac{c^3}{G\hbar }\frac{A}{4}} \right) = \frac{1}{\exp S_0 }\exp
\left( {\frac{4\pi \,G}{\hbar c}m^2} \right) \\
 \quad \quad = \frac{1}{\exp S_0 }\exp \left( {\frac{\pi }{G\hbar c}H_0^4
\Omega _m^2 R_h^6 } \right)\,, \\
 \end{array}
\end{equation}

\noindent
where $H_0 $ is the present-day Hubble constant, and $R_h $ the size of the
horizon. Using the case of an event horizon which is fixed by the magnitude
of the cosmological constant only (e.g., Ref. 26),

\begin{equation}
\label{eq2}
R_h = \frac{c}{H_0 \sqrt {\Omega _\Lambda } },
\end{equation}

\noindent
we obtain the following remarkable expression (for the flat $\Omega _m +
\Omega _\Lambda = 1$ universe):

\begin{equation}
\label{eq3}
N_{\min } \sim \exp \left( {\frac{\pi \,c^5}{G\hbar }\frac{\Omega _m^2
}{\Omega _\Lambda ^3 }\frac{1}{H_0^2 } - S_0 } \right) = \exp {\kern 1pt}
\left[ {\frac{\pi \,c^5}{G\hbar }\frac{\Omega _m^2 }{\left( {1 - \Omega _m }
\right)^3}\frac{1}{H_0^2 } - S_0 } \right].
\end{equation}

We notice the appearance of all major constants of nature (including the
``silent'' Boltzmann constant, which is omitted since we are working in
``natural'' units!) in this formula, with the exception of the elementary
charge, which is reasonable since we are dealing with the standard
electrically neutral universe. In addition, the total entropy of our
cosmological domain, $S_{0}$, appears and it represents, in a sense, the
outcome of all and every process which has actually taken place since the
beginning of time!

How big is the realistic entropy $S_{0}$? The conventional answer is simple:
the entropy is by far dominated by the photons of the cosmic microwave
background, whose specific entropy (``entropy-per-baryon'') is a well-known
dimensionless number (e.g., Ref. 21):

\begin{equation}
\label{eq4}
s_{CMB} = \left( {\frac{n_\gamma }{n_B }} \right)_0 \approx 10^8,
\end{equation}

\noindent
where $n_\gamma $ and $n_B $ are number densities of photons and baryons
respectively. Taking the standard estimate that the total number of baryons
within horizon is $\sim $10$^{80 - 81}$, we may be certain that $S_{0}$ is
not larger than 10$^{90}$ (in natural units as well).

The common numerical values of the cosmological parameters $\Omega
_m $ ($ \approx $ 0.3) and $H_0 $ ($ \approx $ 60 km s$^{ - 1}$
Mpc$^{ - 1})$ inserted in (3) give us a stupendous double
exponential

\begin{equation}
\label{eq5}
N_{\min } \sim \exp \left( {1.9\times 10^{121}}
\right). \quad {\rm (!!!)}
\end{equation}

At least that many domains in the multiverse are needed to account
for the observed asymmetry in this manner.$^{(27)}$ (This is
easily generalized to the case of charged or rotating universe
characterized by some other set of parameters, but their exact
values make almost no difference when numbers of such magnitude
are involved.) This is the price one must pay for embedding the
atypical initial conditions into a wider manifold (see below, \S
4). Of course, the total number of domains may be infinite, in
which case the conclusions of Ellis and Brundrit$^{(28)}$ will
apply, and any worry about the ``specialty'' of our initial
conditions is immediately discarded. On the truly global
scale---i.e.\ in the multiverse---there is no thermodynamical
asymmetry, no arrow of time.

Note that the end result of both this and the two proposals Price
describes---given by the parsing schemes above---is the same: it
is an \textit{observed asymmetry}. However, the attribute seems
superfluous in both Acausal-Particular and (especially)
Causal-General approaches. It has no function at all in either
approach. Only in the anthropic approach advocated here it does
receive a proper place in the \textit{core\/} of the perceived
explanandum. Is ``observed'' in this context the same as
``objective'' or it is not? By equivocating on this, Price
attaches a strongly realist (indeed, essentialist) character to
the two approaches he favors. Boltzmann and Schuetz, on the
contrary,adopt more cautious stance and employ the attribute
``observed'' in its true and literal meaning (i.e.\ observed by
intelligent observers possessing specific capacities, epistemic
and otherwise).

In other words, a symbolic way of \textit{roughly} representing the relationship of the
present approach to the Boltzmann-Schuetz hypothesis is the following
scheme:

\vspace{0.4cm}

Acausal-Anthropic approach = (Boltzmann measure + anthropic selection
effect) +

+ the multiverse =

= Boltzmann-Schuetz picture + the multiverse.

\vspace{0.4cm}

(It is necessary to qualify this as ``roughly'', since the cosmological data
of the XX century has been fully taken into account in the present approach,
data which Boltzmann was, naturally, completely ignorant about.) The
multiverse or world ensemble has by now become quite a familiar term to
cosmologists, as well as to at least some philosophers. Thus, Bostrom
cogently writes:$^{(29)}$

\begin{quotation}

...The meaningfulness of the [world] ensemble hypothesis is not much in
question. Only those subscribing to a very strict verificationist theory of
meaning would deny that it is possible that the world might contain a large
set of causally fairly disconnected spacetime regions with varying physical
parameters. And even the most hardcore verificationist would be willing to
consider at least those ensemble theories according to which other universes
are in principle physically accessible from our own universe. (Such ensemble
theories have been proposed, although they represent only a special case of
the general idea.)

\end{quotation}

The anthropic selection effect is nothing particularly new either. It was
the great French physicist, mathematician and philosopher Henry Poincar\'{e}
who first noted that the functioning of an intelligent mind would have been
impossible in an entropy-decreasing universe.$^{(30)}$ Later it was
elaborated by Norbert Wiener in his celebrated \textit{Cybernetics} (Ref. 31, p. 35):

\begin{quotation}

Indeed, it is a very interesting intellectual experiment to make the fantasy
of an intelligent being whose time should run the other way to our own. To
such a being, all communication with us would be impossible. Any signal he
might send would reach us with a logical stream of consequents from his
point of view, antecedents from ours. These antecedents would already be in
our experience, and would have served to us as the natural explanation of
his signal, without presupposing an intelligent being to have sent it. If he
drew us a square, we should see the remains of his figure as its precursors,
and it would seem to be the curious crystallization---always perfectly
explainable---of these remains. Its meaning would seem to be as fortuitous
as the faces we read into mountains and cliffs. The drawing of the square
would appear to us as a catastrophe---sudden indeed, but explainable by
natural laws---by which that square would cease to exist. Our counterpart
would have exactly similar ideas concerning us. \textit{Within any world with which we can communicate, the direction of time is uniform}.
\end{quotation}

What the father of modern computer science had in mind, is that the
existence of an entropy gradient of particular size in a sufficiently large
region of space is a necessary precondition for the cognitive operations of
intelligent life as we know it. In order to immediately preempt any
misgivings about the alleged anthropocentrism which undeservedly follow
practically any anthropic argument, it should be explicitly stated that this
property does not have any particular association with \textit{homo sapiens} (except the trivial
and temporary one that we are the only certain example of intelligent
observers known so far; that circumstance is likely to change soon, as a
result of either SETI or AI efforts).

Of course, the premiss that intelligent observers select
particular entropic behavior (and thus entail a temporal
asymmetry) should not be taken for granted---and here we come to
the crux of the explanatory task ahead. Instead of searching for
strange, hitherto never seen time-asymmetric microprocesses (as in
the Causal-General approach), or attempt to find a new law
applying to global singularities (as in the Acausal-Particular
approach), here we would like to investigate and ascertain
\textit{why} intelligent observers are dependent on the entropy
gradient to function. This, obviously, brings explanatory tasks
into the realm of information theory, but also to the disciplines
like complexity theory, cognitive sciences, and neurosciences. But
this does not mean these are not physical questions! An awareness
that the link between information theory and various physical
theories is the \textit{centre piece} \textit{of any attempt at
understanding nature} has been rapidly growing over several
decades. Since the brilliant book by Brillouen$^{(32)}$,
physicists are gradually getting accustomed to the idea, (e.g.,
Refs. 33-35), and this struck cord with philosophers too (for a
particularly nice example, see Ref. 36). On the other hand, the
working philosophy of computationalism has become established as
the basis of the bulk of cognitive sciences$^{(37)}$. In all
quarters one may nowadays hear scholarly debates on ``physical
reductionism'', ``monism'', ``physicalism'' and many other labels
which pertain to essentially the same thing: that cognition (and
the various phenomena associated with it, including the apparent
temporal asymmetry) is, in principle, explicable in physical
terms. That we are still far from such an explanation, is
certainly unnecessary to elaborate upon. Thus, the project
suggested here is neither non-physical nor easy!\footnote{ As an
aside, let us note that the issue of the psychological arrow of
time has been lightly and casually dismissed by most writers of
the temporal (a)symmetry as a (matter-of-course) consequence of
the thermodynamical arrow (e.g. Ref. 38). This uncritical behavior
has been criticized in the influential paper of John
Earman$^{(39)}$, and we might add only that, given the
contemporary state of the issue, the critique has been too mild!
There is no single piece of hard evidence produced so far that the
psychological arrow can be reduced in such way to the
thermodynamical one; one of the priority tasks within the proposed
Acausal-Anthropic approach would be to clarify this difficult
issue from the point of view of the information theory and
cognitive sciences.}

We notice that this still represents the one-asymmetry approach, a
sort of a cousin to the Acausal-Particular approach. However, the
asymmetry is located at a different place from the one in the
Acausal-Particular approach favored by Price. \textit{Very
roughly\/} speaking, we need information theory and cognitive
science rather than quantum gravity or even the ``Theory of
Everything''---which presumably, although Price remains silent on
the issue, determines the nature of the initial low-entropy
conditions necessary for operation of the Acausal-Particular
mechanism. The fact that we seem to know more about the former
than about the latter (especially since quantum cosmology, through
its multiverse theories, demonstrates that the true ``Theory of
Everything'' may not exist at all; cf.\ Refs. 40-42) is another
advantage of the Acausal-Anthropic approach.

A possible objection against this explanatory project may be
formulated as follows. How did the multiverse come into being? If
it came into being by, say, a particular form of spontaneous
symmetry breaking, would not that count as manifestly
unidirectional process in time? In other words, it may look as if
the concept of the multiverse itself \textit{presumes} an arrow of
time.\footnote{ Special thanks are due to one of the anonymous
referees for bringing the author's attention to this interesting
point.} This certainly contradicts what we have said about
Boltzmann's statistical measure and the totality of cosmological
domains being in perpetual thermodynamic equilibrium. Two
recourses are available for defense of the Acausal-Anthropic
approach, both belonging to the core of quantum cosmology (and the
accompanying ontology) and each being worthy a lengthy discourse
in its own right. First, we may argue that the multiverse, being
the grand stage for everything that happens, holds a privileged
position and is not further explicable. So, any theory pertaining
to explain the multiverse itself would be superfluous, since the
multiverse is a ``brute fact''. This would be the most
straightforward generalization of the original Boltzmann-Schuetz
idea (cf. Chapter 2 of Ref. 43). A more interesting defensive
strategy would be to argue that the critique actually restricts a
range of possible multiverses to those which are, in Linde's terms
``eternal, self-reproducing fractals.'' Its eternal nature
obviates the need for explaining its origination, and its
self-reproducing feature enables what philosophers call
``explanatory self-subsumption.'' If the principle necessarily has
the feature it speaks of, then it necessarily will apply to itself
(cf.\ Ref.\ 44). Individual domains---``universes''---are created
and destroyed (for instance, through return to the Planck
space-time foam upon recollapse), but the multiverse itself stays
completely time-reversible. Of course, further elaboration of
these ideas is certainly necessary, but the objection does not
seem to be fatal at the moment.

So, what are the additional benefits of this project, beyond a
novel look at the explanandum of thermodynamical asymmetry? We
have already seen some of them: dropping the \textit{ceteris
paribus\/} clause, for instance. We are now free to state that
entropy is always high in by far the predominant part of
everything that exists. Thus we avoid ``a surprising consequence
of the one-asymmetry view:''$^{(9)}$ the fact that Boltzmann's
statistics---being based upon temporally symmetric
probabilities---implies high entropy in the past as well as in the
future. And we avoid it in a natural and simple manner, which
Boltzmann has endorsed himself!

However, the greatest advantage comes from the possibility of connecting to
cosmology, and especially contemporary currents in quantum cosmology. This
is achieved without much further effort which the Acausal-Particular
approach necessitates. From Bruno, who argued for innumerable worlds by
using a specific form of the principle of plenitude, to Hume and his
``innumerable worlds botched and bungled,'' to Linde, Vilenkin and other
modern cosmologists, as well as some respected contemporary
philosophers,$^{(43,45)}$ people by and large did not take this issue
lightly and casually. There are several different approaches which all lead
to the conclusion that what we perceive as our cosmological domain is just a
minuscule fraction of everything that exists (cf. Refs. 40, 46).

A brief historical aside seems due at the end of this section. The
Acausal-Anthropic view in the modern (post-Boltzmann) sense was
first formulated by Clutton-Brock$^{(47)}$. As the story often
goes, this highly unorthodox contribution was published in an
astronomical journal of rather modest circulation and went almost
unnoticed. Clutton-Brock explicitly considered the distribution of
the initial entropy-per-baryon over the set of existing
\textit{worlds}, and concluded that only a relatively narrow range
of this quantity permits the formation of the complex structures
we perceive and, presumably, the formation of life
itself.\footnote{ It should be noted that Clutton-Brock was not
dogmatic on this; in Ref. 48 he proposes a version of the Gold
model close to the one Price favors. } The importance of his
results has, however, been slightly obscured by some unnecessary
assumptions, as well as by confusion over the two types of the
multiverse that have been proposed (see footnote 2). Similar
analysis for the parameter $Q$, describing the amplitude of cosmic
microwave background fluctuations, has been given only recently by
Tegmark and Rees$^{(49)}$, who discuss the anthropic selection
effects plausibly underlying the magnitude of the anisotropies of
the early universe, as detected recently with \textit{COBE}. This
point is directly related to the problem of low initial entropy,
since these authors correctly identify the amplitude of these
fluctuations in microwave background radiation with the amplitude
of gravitational potential fluctuations in the early universe when
they enter the horizon. The observed fact that this number is of
the order of $Q \quad \sim $ 10$^{ - 5}$ cannot be derived from
known physical theories, and---as Tegmark and Rees emphasize at
the very beginning of their paper---one may either wait for some
future fundamental theory from which $Q $could be computed or take
the option (supported independently by various inflationary
scenarios) that it is effectively a random number drawn from some
wide distribution whose observed values will be constrained by the
anthropic selection. What they persuasively demonstrate in the
rest of their paper is that such constraints are effective in
keeping the expected \textit{observed }value in the approximate
interval $10^{ - 6} \le  Q \le  10^{ - 4}$.


\Section{3. IS BOLTZMANN-SCHUETZ INTRINSICALLY UNTENABLE?}

\noindent
What problems may the Acausal-Anthropic view face? There
are two objections to the \textit{Boltzmann-Schuetz version} of
this view, both of which have surfaced from time to time in the
literature. We shall denote each by using the names of the two
famous 20th century physicists who elucidated them. Before we
investigate their status, two issues should be kept in mind.

(I) Strictly speaking, the Acausal-Anthropic view we propose does
not need to answer these objections, since the multiverse is not
necessarily the same topologically connected entity to which these
objection refer. However, we shall show the weakness of these
arguments even against the classical Boltzmann-Schuetz picture, in
order to re-display the one-sidedness of Price's account, who
almost casually dismisses this proposition.\footnote{ Telling is
his locution that anthropic selection suggested by Boltzmann makes
``this option \underline {less unappealing} than it initially
seems.'' (Ref.\ 9, \S 3.4---underlined by M. M. \'C.) } An
important motivation for doing this is that we are still very much
in the dark about the detailed physical nature of the multiverse
(and in particular, the issue of its causal structure). But,
obviously, if we weaken the arguments against the ``one- domain''
anthropic picture (i.e. the classical Boltzmann-Schuetz view), the
multiverse view may just benefit.

(II) \textit{Any reasonable defense} of the Boltzmann-Schuetz
picture must be qualified, since the progress of modern cosmology
has made obsolete practically all ideas about the universe on the
largest scales which were current at the end of the 19th
century.\footnote{This point is probably the content of Price's
remark (Ref. 8, \S 1) that ``this strategy [Boltzmann-Schuetz],
never entirely satisfactory, seems to have been decisively
overturned by recent work in cosmology.''} Thus, this theory is
certainly indefensible when set against the objection that the
empirically established age of (our) universe is much shorter than
the timescale required for large entropy fluctuations. Many
other---mainly astrophysical--- objections, related to
inhomogeneities in the distribution of matter, the universal
photon heat bath, entropy production in stars and black holes,
quantum properties of matter, the existence of horizons, etc.,
could also be stated if the Boltzmann-Schuetz picture is to be
taken literally nowadays. After all, that is the main reason why
we have adjoined the multiverse concept to this picture in
presenting the Acausal-Anthropic view above. In contrast, in the
rest of this section we shall question the \textit{intrinsic\/}
tenability of the Boltzmann-Schuetz picture, without entering too
much into the empirical considerations stemming from perceived
astrophysics. The point is to show that the classical objections
are significantly overrated, and that by embedding this picture
into contemporary multiverse theories we can gain much more than
we can lose. In other words, the death of the
anthropic-fluctuation picture has been somewhat prematurely
proclaimed.

That said, let us consider the argument Price and other critics put forward
against the Boltzmann-Shuetz view. We have:

1. \textbf{Feynman's argument} (FA): why is the size of low-entropy
fluctuation so much larger than is necessary for the emergence of
intelligent observers on Earth?$^{(50)}$

2. \textbf{von Weizs\"{a}cker's argument} (vWA): How could we believe the
information of a low entropy past, when a smaller (and \textit{eo ipso} likelier)
fluctuation is sufficient to produce such (false or misleading)
evidence?$^{(51)}$

Let us consider FA first. There is an obvious way out (first hinted at by
Schelling in the early 19th century!): simply deny the antecedent---namely
the assertion that less (low entropy) space is required for the emergence of
intelligent observers than we perceive. This assertion is by no means
obvious, since we still know too little about the physical and chemical
preconditions for the origin of life, not to mention intelligence. Suppose,
for instance, that the conventional Oparin-Haldane (``warm little pond'')
hypothesis for the origin of life (e.g., Ref. 52) is unsupported or wrong
(as many researchers have claimed; e.g., Refs. 53-55). The most serious
scientific alternative to the received view is the panspermia hypothesis:
the view that life was formed at some other place in the universe and has
been brought to Earth via either terrestrial encounters with interstellar
dust, clouds or comets or even the intentional action of an advanced
galactic civilization$^{(56 - 58)}$. It is immediately clear that the \textit{size}
requirements of these two views (Oparin-Haldane vs. panspermia) are
significantly different. If the \textit{a priori }probability of a spontaneous assembly of
complex molecules necessary for the creation of life is of the order of
10$^{\mbox{--}3000}$ (or even less, as has been suggested in the references
critical of the received view), then we are entitled to insist on a much
larger spatiotemporal volume than the conventional ``warm little pond'' on a
single planet. Such a larger volume is necessary even if we know, for
instance, of only one single isolated human observer in his cubicle
somewhere on Earth.

An FA proponent could retort that panspermia is a highly speculative idea.
Of course, it certainly is to a degree because science lacks the universal
yardstick of ``speculativeness''. The received view is also quite
speculative.\footnote{ And the Gold-type recollapsing universe suggested by
Price in the context of his favorite Acausal-Particular approach is far more
so.} But another point is central here. Not wishing to enter into a
biological discussion about which of these options is more likely to account
for the totality of empirical data, it seems clear that we should not
\textit{presume} any one of them to be correct when evaluating proposals for the origin of
the entropy gradient---and that is exactly what FA does. It puts the
explanatory cart in front of the horse.

Weakness of FA is graphically manifested if we try to cast it into a
somewhat more rigorous form. Let us consider a causal chain of $n$ events
e$_{1}$, e$_{2}$, ..., e$_{n}$ leading to the emergence of intelligent life
capable of reading Feynman and Price:



\begin{equation}
e_{1} \;  \Rightarrow  e_{2} \;  \Rightarrow  ...
\Rightarrow  e_{n} = {\rm existence\ of\ sophisticated \ int.\ obs.\ on
Earth,}
\end{equation}


(for the sake of the argument, we shall assume that $n$ is a
finite natural number, and that there is a well-defined first
member of the sequence). Examples of links in this chain are the
formation of oxygen in the primitive Earth's atmosphere, or the
synthesis of $^{12}$C in stars through the triple-$\alpha$
reaction.$^{(55,59)}$ Now, we can, in principle, assign a volume
V(e$_{i})$ to all links in the chain as

V(e$_{i}) \quad  \equiv $ minimal spatial volume necessary for e$_{i}$ to occur,

\noindent
which is intuitively clear (one cannot expect galaxy formation to occur
within 3.6 m$^{3}$, etc.). FA can therefore be translated as a statement

\begin{equation}
(\forall i) V(e_{i}) \ll (R_{H})^{3}, 
\end{equation}

\noindent where (R$_{H})^{3}$ is any prototype of the
representative part of the observable universe, which may be
visualized as a cube with sides equal to the Hubble radius (or
less, depending on our assessment of the high-redshift
astrophysical data). Is (7) the truth so self-evident that anybody
(with Boltzmann and Schuetz) who doubts it may be casually
dismissed as ``ridiculous''? Not likely. What is max[V(e$_{i})$]?
Not so obvious either. As Barrow and Tipler$^{(21)}$ correctly
note, if the process of cosmological structure formation is a
necessary part of the causal chain (6), we have every reason to
believe the contrary proposition, namely (e.g., Ref.\ 60)

\begin{equation}
(\exists j) \; V(e_{j}) \quad  \approx  (R_{H})^{3}.
\end{equation}

Of course, this is part of the modern, evolutionist view; in the
time of Boltzmann and Schuetz, everything was considerably easier,
since the Hubble expansion and consequent developments were more
than thirty years in the future. Nowadays, one may justifiably ask
whether R$_{H}$ in (7) and (8) should be temporally indexed, so
that it pertains to the relevant cosmological scale \textit{at
epoch of e}$_{i}$. This makes FA even less persuasive: if some
perturbation, phase transition, or any other part of the quantum
cosmological lore is truly necessary for our emergence as
intelligent beings, then we have every right to conclude that the
entire observable universe is an ``all or nothing''
matter.$^{(40)}$

Another way to visualize this option is to imagine that there is a
minimum spatial volume (or ``cell'') which is necessary in order
for intelligence to arise at a single point within such volume.
Each point in spacetime is either within a cell or it is not. If
you are within a cell, then you perceive stars, galaxies,
etc.---the entire familiar low-entropy universe. If you are not,
then you perceive something of high entropy, presumably lots of
black holes---but you cannot be there. How big is the volume of
the smallest cell? Feynman estimates it to be much smaller than
the observed universe (essentially the Hubble sphere); still there
is no argument except crude intuition to that effect.

Price himself testifies that vWA is not an overwhelmingly
persuasive objection: when treating globally symmetric cosmologies
(\`{a} la Gold) he explicitly dismisses vWA---and in a footnote at
that, obviously not deeming it necessary to invoke an elaborate
explanation. The same issue happens in his discourse on the
epistemology of the relationship between randomness and high
entropy (footnotes 6 and 7 in Ref.\ 9): ``Again, I am ignoring von
Weizs\"{a}cker's sceptical difficulties about inference to the
past.'' Remarks such as this provoke a simple question which one
is fully entitled to ask of Price: can von Weizs\"{a}cker's view
be ignored or not, in the final analysis? Because, if it is not
ignored when disputing the anthropic approach to the explanatory
project (and thus, indirectly, disputing what we have called the
Acausal-Anthropic approach), and simultaneously it is conveniently
ignored in the elaboration of the Acausal-Particular approach,
then it seems clear that we are operating a double standard.

However, in order to be fair toward vWA, we should not construe
this double standard as a counter-argument. A true
counter-argument is to analyze things further: what exactly in vWA
is \textit{problematic\/} for the Boltzmann-Schuetz picture? Of
course, vWA ridicules our epistemic capacities---but that is
\textit{non sequitur}. Remember that Boltzmann and Schuetz arrived
at their view exactly starting from a high-entropy \textit{default
condition}, which does not call for an explanation (as Price is
entirely correct in proclaiming). They have not relied in deriving
their hypothesis on any one specific piece of empirical data, in
particular any empirical data telling of \textit{past
conditions\/} of the universe. Thus, their hypothesis cannot be
refuted by accepting that such-and-such empirical data which tells
us about a low-entropy past is either ``simulated'' or invalid.
vWA may conflict with propositions embodying the weak anthropic
principle (that the values of physical and cosmological parameters
are constrained through the requirement of existence of the
carbon-based life, sufficiently stable conditions, etc.), but that
is only because anthropic theoreticians have not bothered to cope
with the case of simulated evidence. Another issue is that
``simulation'' cannot be as simple as it is tacitly assumed in von
Weizs\"{a}cker's or Price's account; the fact that we observe
coherent information flow from the past to the present (as, for
instance, in the Collins' postulate of uniform thermal histories;
Ref.\ 61) seems just a coincidence on this account. This is
consequence of the fact that information-theoretical background of
this possibility has not been investigated in detail so far. The
status of vWA, thus, is that it shows a bizarre and unexpected
side of the Boltzmann-Schuetz view, but does not reveal true
contradiction or incoherency.

Also, one is prompted to ask: how exactly do we prove that
``simulation'' is cheaper (in entropy terms) than ``reality'' in
all (or just most of) cases? Again, the opponent of
Boltzmann-Schuetz (von Weizs\"{a}cker this time) offers just a
hunch and ``common sense'' intuitions. The question is admittedly
difficult and depends again on the information-theoretical
uncertainties mentioned above. There is at least one serious study
claiming the contrary, the one of Tegmark;$^{(35)}$ or, more
precisely, it claims that in cases of an extremely high degree of
symmetry (with help of Everett's quantum theory and the
environmental decoherence), the algorithmic information content of
the entire observable universe may be very small, essentially the
same one as in the earliest moments of its existence. In Tegmark's
view (especially appealing to cognitive scientists and information
theory researchers), the whole is much simpler than its parts, and
almost everything we perceive is a sort of illusion. In
particular, the apparent complexity of everything around us---not
just evidence of the past---is illusory (or "simulated")! We may
find Tegmark wrong on several issues, but the lesson of his work
is that our current methods (in the theory of complexity and
related disciplines) are far from being entirely reliable and
disputable in the matter of the entropy cost of perceived pieces
of reality.

Let us note that the two arguments discussed in this section are
not even strictly co-tenable. FA cannot be coherently formulated
if we assume that vWA is correct (since what then is the size
which matters? how can we define it, since our knowledge on the
properties of life itself---not to mention the standards of
measurement---is based upon simulated evidence?). Conversely, on
FA it would be idle speculation to ask ``what is the real nature
of the universe outside of our small confines?'', as vWA would
prompt us to. (One may play off one against the other, claiming
that, for instance, an intelligence-bearing cell of necessary low
entropy is the size of our Galaxy, and all extragalactic
information is simulated. There have been similar suggestions in
the older cosmological literature and in a different context,
notably so-called McCrea's uncertainties; see Ref.\ 62)


\Section{4. EMBEDDING THE ATYPICAL INITIAL CONDITIONS}

\noindent It has been well-known for quite some time that there
are three basic approaches to answering the question ``Why the
initial conditions of the universe were such-and-such?'' The first
rejects the validity of the question. It may be motivated either
by positivistic refusal to discuss issues forever closed to any
form of direct verification or theological reasons (all standard
theistic accounts of the creation belong to this option). This
rather nihilistic option should only be entertained as a truly
last resort, a counsel of despair. Of the other two,
one---causal---entails the idea that there is a law-like reason
(presumably to be derived from the future ``Theory of
Everything'') for the atypical or surprising structure of the
early universe. In other words, an enormous amount of information,
necessary for the description of the atypical initial conditions,
can be encoded in some new law(s) of nature and consequent
law-like correlations of various matter and vacuum fields. The
other---anthropic---option avoids giving a specific description by
embedding those conditions into a sufficiently symmetric
background. Again, stated in terms of information, the same long
description of what we perceive as atypical initial conditions
arises---as so often in physics!---from the process of
\textit{symmetry breaking}. The overall description is simple
enough, and may be reduced (in the extreme case) to a rule similar
to ``All possible combinations of initial conditions exist.'' That
such a high degree of symmetry can indeed completely reproduce the
situation in our particular domain becomes then an immediate
consequence (cf.\ Refs.\ 35, 63).\footnote{ A brief literary
analogy may be helpful here. In the beautiful story of ``The
Library of Babel'', Jorge Luis Borges has described a world
consisting of a huge library in which piles of seemingly
completely random books are stored.$^{(64)}$ The nature and
content of each book look extremely puzzling to the inhabitants of
that world, most of whom have never encountered a book with a
single meaningful line. Gradually, they reveal the truth: although
each book \textit{per se\/} requires lots of information to be
described---``long description''---the entire content of the
Library is extraordinary simple: it is completely described by the
following proposition: \textit{All variations of letters and
punctuation marks exist in the Library}. Thus, although lost works
of Plato or Tacitus are \textit{certainly\/} located somewhere in
the Library (and are presumably \textit{unique\/}), the wealth of
information they contain from a human perspective is completely
lost among myriads of less and less similar copies of the supposed
original, and myriads of completely worthless ``chaotic'' volumes.
Similarly with the multiverse. A book containing perfect Socratic
dialogues stands in the same relation towards the Library as a
domain containing intelligent observers relates to everything that
exists. Note that the number of books in the Library is, as an
anonymous genius discovers in the story, still finite, although
incredibly huge, and thus the number of meaningful books is
\textit{not\/} of measure zero in the entire set. Similar is the
situation with the domains of the multiverse, as discussed above.
In the infinite case, the number of those allowing for emergence
of intelligent observers is probably of measure zero.}

In a sense, the appeal of the anthropic selection effect is, figuratively
speaking, to kill two birds with one stone: to set up a project for the
explanation of common thermodynamical asymmetry and to obtain the cheapest
possible (in both physical and epistemological terms) explanation of the
initial conditions, in particular vis-\`{a}-vis cosmological fine tunings.
Appeal to the multiverse enables us to ``cleanse the door of perception''
and overcome the narrow coffines of our specially restricted viewpoint.

Price's suggestion (Ref.\ 8, \S 3.3) that the anthropic selection
effect(s) may be necessary only for an explanation of the initial
conditions of our domain---and even then it is obviously less
interesting option for him, since it can be conveniently
``ignored''---certainly does not do justice to the scope and
ingenuity of anthropic reasoning. His preferred option, the
existence of a law-like reason for low entropy is, in effect, a
strange retreat from the \textit{acausal\/} behavior of
thermodynamical systems correctly emphasized at the present time.
Granted that there may be interesting reasons for exploring the
law-like possibility, we are still entitled to ask: why should
one, after rejecting the law-like behavior of matter in the
present epoch, ask for a new law-like behavior in epochs long
gone---say at Planck time---to explain the very same thing?
Especially if a plausible (and rather more ``Boltzmannian'')
anthropic alternative is at hand?

\Section{5. CONCLUSIONS}

\noindent
We conclude that a generalized Boltzmann-Schuetz or
Acausal-Anthropic approach may be able to account for the
perceived thermodynamical asymmetry at least in the same degree as
the other approaches explicated in the recent literature. Neither
of the two conventional arguments is really decisive and
sufficient to reject the Boltzmann-Schuetz anthropic fluctuation
picture. \textit{A fortiori}, these are either irrelevant or
insufficient to reject the Acausal-Anthropic view proposed here.
(Which makes this approach largely insensitive as to the
speculative issue of the exact physical nature of the domains
within the multiverse.)

An enormous additional benefit comes, of course, from being able to account
for other anthropic coincidences (or 'fine-tunings') within the same
conceptual framework received from the rapidly developing ideas of quantum
cosmology. Further advantages include a connection with the latest thinking
in quantum cosmology (incorporating the idea of the multiverse), dropping of
the \textit{ceteris paribus} clause in specification of the default thermodynamical condition,
better accounting for thermodynamical counterfactuals and obviation of the
necessity to double-deal with cosmological data. At the same time, many of
the virtues of the Acausal-Particular approach are retained in the
Acausal-Anthropic picture.

Future behavior of the universe is rapidly becoming a recognized and
legitimate target for ``everyday'' scientific work, and less and less an
arena for wild speculation and ``grand principles.''$^{(65)}$ The nascent
discipline of physical eschatology (e.g., Ref. 66) has already reached many
interesting results, and it is highly misleading to present contemporary
astrophysicists as ignorant about the subject as their colleagues in the
time of Boltzmann or Haldane or Gold. According to unequivocal conclusion
drawn from these empirical developments, the asymptotic final state of our
cosmological domain (or ``universe'') will be one of extremely diluted
matter and extremely high entropy (at least as long as one keeps possible
intentional actions of advanced intelligent communities out of the picture).

\vspace{1cm}

\noindent
\textbf{Acknowledgements.} Any possible merits of this
work are due to Irena Dikli\'c, whose curiosity, tenderness and
kind support have presented a constant source of inspiration and
encouragement during the work on this project; its completion
would be utterly impossible without her. It is superfluous to
emphasize that the responsibility for any shortcomings in the
present manuscript rests solely and on the author. Useful
discussions on the subject with Zoran \v{Z}ivkovi\'c, Nick
Bostrom, Petar Gruji\'c, Slobodan Popovi\'c, and Nikola
Milutinovi\'c have led to the significant improvements in both the
content and the presentation. In addition, the author uses this
opportunity to thanks Vesna Milo\v{s}evi\'c-Zdjelar, Maja
Bulatovi\'c, Milan Bogosavljevi\'c, Nick Bostrom, Vladan \v
Celebonovi\'c and the Embassy of Netherlands in Belgrade for their
kind help in obtaining some of the references. Four anonymous
referees are acknowledged for suggestions useful in improving the
overall quality of the manuscript. Invaluable technical help has
been received from Srdjan Samurovi\'c and Alan Robertson.

\vspace{1cm}

\def\refname{{\bf REFERENCES}}

\end{document}